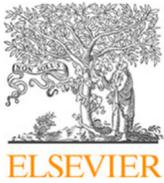
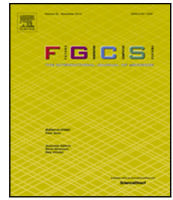

Review article

# Secure integration of 5G in industrial networks: State of the art, challenges and opportunities

Sotiris Michaelides [a],[*], Stefan Lenz [a], Thomas Vogt [a], Martin Henze [a],[b]

[a] *Security and Privacy in Industrial Cooperation, RWTH Aachen University, Im Süsterfeld 9, 52072 Aachen, Germany*
[b] *Cyber Analysis & Defense, Fraunhofer FKIE, Fraunhoferstraße 20, 53343 Wachtberg, Germany*

## ARTICLE INFO

## ABSTRACT

The industrial landscape is undergoing a significant transformation, moving away from traditional wired fieldbus networks to cutting-edge 5G mobile networks. This transition, extending from local applications to company-wide use and spanning multiple factories, is driven by the promise of low-latency communication and seamless connectivity for various devices in industrial settings. However, besides these tremendous benefits, the integration of 5G as the communication infrastructure in industrial networks introduces a new set of risks and threats to the security of industrial systems. The inherent complexity of 5G systems poses unique challenges for ensuring a secure integration, surpassing those encountered with any technology previously utilized in industrial networks. Most importantly, the distinct characteristics of industrial networks, such as real-time operation, required safety guarantees, and high availability requirements, further complicate this task. As the industrial transition from wired to wireless networks is a relatively new concept, a lack of guidance and recommendations on securely integrating 5G renders many industrial systems vulnerable and exposed to threats associated with 5G. To address this situation, in this paper, we summarize the state-of-the-art and derive a set of recommendations for the secure integration of 5G into industrial networks based on a thorough analysis of the research landscape. Furthermore, we identify opportunities to utilize 5G to enhance security and indicate remaining challenges, identifying future academic directions.

## Contents

* Corresponding author.
   *E-mail addresses:* michaelides@spice.rwth-aachen.de (S. Michaelides), lenz@spice.rwth-aachen.de (S. Lenz), vogt@spice.rwth-aachen.de (T. Vogt),
henze@spice.rwth-aachen.de (M. Henze).

## 1. Introduction

Industrial Control Systems (ICSs) traditionally utilize wired technologies such as Ethernet and a variety of protocols such as Modbus, PROFINET, or EtherNet/IP, to realize the communication between the different components [1]. However, the digital transformation of production, driven by trends such as Industry 4.0 or the Industrial Internet of Things (IIoT) [2,3], introduces new demands for performance and modernization that the next generation of industrial networks must meet. In contrast to previous wireless technologies such as Long Term Evolution (LTE) and Wi-Fi, 5G promises to fulfill the requirements for availability, low latency, reliability, the interconnection of numerous devices, and the modernization of the networks [4]. Additionally, 5G enables mobility and introduces new use cases such as remote-real-time control, and low latency access to cloud resources that improve the functionality and automation of the ICS [5].

However, the integration of 5G as the communication infrastructure in industrial networks raises serious concerns due to the introduction of new security threats, its inherent complexity, and its wireless nature. Considering that, until recently, industrial networks operated on simple, fieldbus-based systems that were isolated from other networks, these concerns are well-justified. Recent examples of cyberattacks in industrial networks, such as STUXNET [6] and the attack on the Ukrainian power grid [7], are proof that the security of industrial networks must be prioritized. These incidents demonstrate that industrial networks are becoming the targets of powerful adversaries and represent a new way of conducting warfare [8]. This becomes even more important when introducing new, and complex components to the ICS, such as 5G. Despite its benefits, 5G also increases the attack surface of the ICS, by introducing new technologies, components and a wireless interface. Therefore, securely deploying and configuring all relevant components of 5G within industrial networks is of utmost importance.

**Related Work.** Over the years, extensive research has been conducted on 5G security, including analyses of real-world 5G implementations, comparisons of the standalone (SA) and non-standalone (NSA) 5G architectures [9], and explorations of private 5G network deployment options with their associated drawbacks and benefits [10–12]. Entities such as the European Agency for Cybersecurity (ENISA) and the Federal Communications Commission (FCC) have investigated the (optional) security controls related to various components of 5G [13,14]. However, these studies often lack consideration for the unique requirements of industrial networks. The 5G Alliance for Connected Industries and Automation (5G-ACIA) has addressed this gap by initiating work on 5G security in industrial networks, emphasizing aspects such as network slicing security and jamming. However, their approach for security mainly treats the 5G network as a closed-box system and thus resorts to the suggestion of using higher-layer security protocols, such as Transport Layer Security (TLS) [15]. In contrast, we argue for the need to comprehensively treat security and especially incorporating the deployment and configuration of the 5G network, especially in industrial networks where critical data, e.g., in the context of Time Sensitive Networking (TSN), is not protected by higher-layer security protocols. In addition to security, 5G-ACIA also identified industry's requirements and explores potential use cases enabled by the integration of 5G [16,17]. Finally, multiple papers examine challenges and opportunities in 5G networks and its associated technologies [18,19]. We differentiate ourselves by conducting our research based on industrial requirements for security, and thus prioritizing availability and safety over confidentiality and integrity.

**Contributions.** In this paper, we perform a comprehensive survey of the state-of-the-art of securely utilizing 5G in an industrial setting, identify open challenges, and highlight opportunities moving forward to realize a secure integration of 5G into industrial networks. To this end, we draw from previous research to summarize the state-of-the-art and to derive a set of recommendations for the secure integration of 5G into industrial networks. Our contributions can be summarized as follows:

- We provide insights into industrial networks, their unique requirements, their growing demand for wireless communication, and 5G as promising solution (Section 2).
- We explore real life 5G deployments to showcase the suitability of 5G in industrial settings (Section 3)
- We summarize the state-of-the-art of securely deploying and configuring industrial 5G networks (Section 4).
- We identify opportunities to further enhance the security of an ICS by utilizing 5G and highlight remaining challenges to drive future research (Sections 5 & 6)

**Impact.** Our overview of the state-of-the-art includes not only the secure deployment of a 5G network but also discusses aspects for a secure configuration and additional security controls. These configurations are particularly relevant in areas where 5G security measures fall short in completely addressing potential threats and vulnerabilities. As such, our work is not only suited for researchers to learn about current research on 5G security in industrial networks but also serves as a guideline for practitioners for a secure deployment and configuration of a 5G system in an industrial network.

## 2. Background: 5G in industrial networks

Industrial networks interconnect various components with different functionalities in an ICS, e.g., to control the physical process and monitor its state. To lay the foundation for our work, we first explain the interaction of these components to control the physical process. Additionally, we identify properties the underlying industrial network must satisfy for the reliable control of the physical process (Section 2.1). Then, we delve into the advantages of wireless technology in general for industrial networks (Section 2.2), provide a concise overview of 5G systems (Section 2.3), and elucidate the role of 5G for carrying industrial data and explain why it is superior to other wireless technologies (Section 2.4).

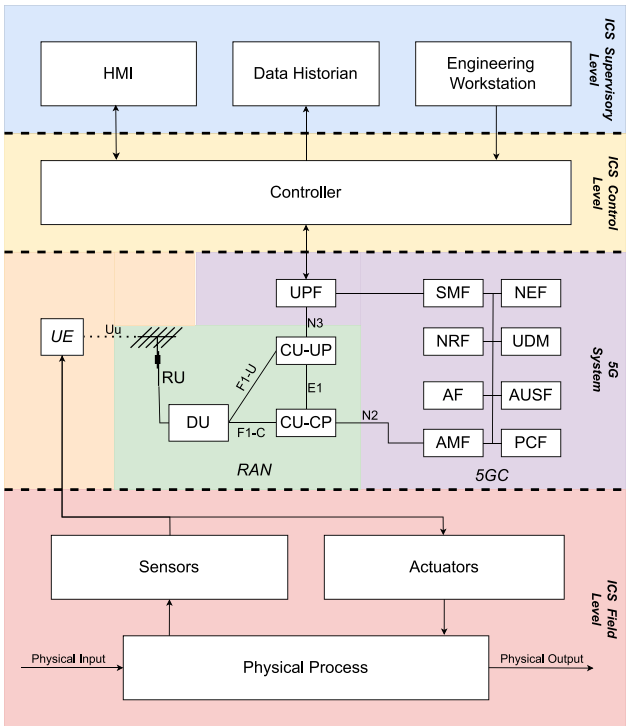

**Fig. 1.** Overview of a 5G-enabled ICS system: 5G replaces the previously-wired connection between sensors/actuators and the controller. Despite its wireless nature, 5G satisfies the requirements of critical Closed-Control Loops, for low latency and real-time operation.

## 2.1. Industrial networks

ICSs are commonly employed to monitor and/or control physical processes in industrial facilities such as nuclear power plants and water treatment facilities. To fulfill its purpose, an ICS, as visualized in Fig. 1, is composed of several units, typically separated into three distinct levels: the field, control, and supervisory levels.

**Layers and Components.** The *field level* is the bottom-most tier, containing the *physical process* itself and its closely surrounding components, namely *sensors* and *actuators*. Sensors are responsible for measuring physical attributes, such as temperature and pressure, while actuators are responsible to perform physical actions, such as opening and closing valves. The *control level* contains the controllers, i.e., logic entities responsible for controlling and coordinating the operation of the ICS. The interaction between these two levels realizes a *closed-control loop* [1]: Sensors sent gathered data of the physical process to the controller. Based on this data, the controller sends commands to the actuators which translate the commands to physical actions. Furthermore, the controller forwards captured data, issued commands and statistics to the topmost layer of the ICS, the supervisory level. The *supervisory level* contains various components responsible for monitoring, logging, and configuration, such as *Human Machine Interfaces (HMIs)*, *Data Historians*, and *Engineering Workstations*.

**Requirements of industrial networks.** In contrast to traditional Information Technology (IT) networks, industrial (or Operational Technology (OT)) networks prioritize availability, real-time operation, low latency, and safety over the typical focus on high bandwidth, confidentiality, and integrity. These distinct requirements led to the development of wired networks focused on performance [20], often lacking basic security controls such as encryption and integrity protection, as these controls introduce additional latency overhead. In many cases, *air-gapping* [1], a method that isolates industrial networks from all other external networks, was the only security measure in place. Recent real-world attacks, such as the *Stuxnet* and the *Night Dragon* [1], proved air-gapping ineffective [21], emphasizing the urgent need to prioritize security rather than relegating it to secondary importance.

Furthermore, the demands of Industry 4.0 and the IIoT for increased automation, extensive cloud-based computational resources, and interconnection of industrial facilities make air-gapping no longer viable. In wireless industrial networks, security controls become even more critical, as propagating radio waves are more challenging to control and secure in comparison to data transmitted through wired connections. For example, anyone within proximity can eavesdrop on or tamper with the traffic [18]. Although wireless communication in industrial networks was previously uncommon, it is now considered a key-component in fulfilling the requirements of the Industry 4.0 and IIoT.

## 2.2. The push for wireless communication

**Benefits of Wireless Networks.** Wireless communication has substantial advantages and enables new use cases in industrial networks [16,17]. Firstly, wireless systems are often deemed more cost-effective than their wired counterparts due to the elimination of extensive cabling and physical infrastructure. This is particularly beneficial for factories with a vast number of IIoT sensors and redundant communication pipelines [22]. Additionally, the elimination of cables enhances scalability and flexibility. Devices can be easily installed without the need for additional hardware, and the wireless network can be accessed from virtually anywhere within its coverage area. Apart from cost reduction, cable elimination also enables mobility and improves the functionality and automation of the factory [12]. With wireless technology, sensors can be attached to rotating or vibrating motors, enabling accurate data collection from dynamic environments, such as measuring the frequency of rotating turbines. In addition, self-driving vehicles, robots, and even people can move without any restrictions or spatial limitations while staying connected to the network, thereby enhancing productivity, collaboration, and enabling new use cases.

**Practical benefits in real-life deployments.** The benefits of wireless deployments have been confirmed by real-world implementations of 5G. For example, Bosch replaced its previously wired and Wi-Fi networks with 5G [23], resulting in annual savings of over €200,000. This transition highlights the significant cost-effectiveness and operational efficiency that 5G can bring to industrial environments. Similarly, a 5G network in a BLISK milling process [24]. By leveraging 5G technology, they were able to introduce motion control into the milling process, significantly enhancing automation. This advancement not only streamlined operations but also led to a reduction in the production cost per BLISK by more than €1000.

Despite their advantages, wireless technologies have historically been underutilized in industrial networks, mainly due to concerns about higher latency. In Section 2.4, we discuss how other popular wireless technologies fail to meet the stringent requirements of industrial applications, positioning 5G as the most viable option for adoption in these environments. However, we precede this discussion with an introduction to 5G and its components in the following.

## 2.3. 5G networks

5G, the latest mobile network generation, was designed to meet three key use cases: *enhanced Mobile Broadband (eMBB)*, *(Ultra-Low Latency Communication (URLLC))*, and *massive Machine-Type Communication (mMTC)*, catering to the demands of end-users and emerging technologies such as the IIoT. While eMBB achieves speeds up to 20 Gbps, URLLC ensures communication with less than 1 ms latency, and mMTC connects 1 million devices per $km^2$, crucial for IIoT and Industry 4.0. As illustrated in Fig. 1, 5G comprises three main components; the *User Equipment (UE)*, the *Radio Access Network (RAN)*, and the 5G Core (5GC) [25]. While 5G can integrate into the ICS anywhere, the Figure presents a typical scenario where 5G serves as the communication infrastructure between control and field level.

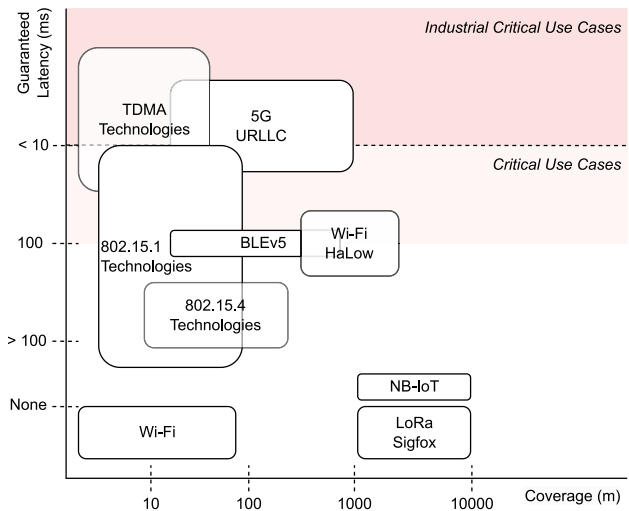

**Fig. 2.** Guaranteed latency comparison between different wireless technologies: 5G is the only wireless technology able to meet industries demand without sacrificing range or bandwidth.
*Source:* Adapted from [27].

The UE, typically a mobile device with a SIM card, accesses the 5GC, a set of interconnected Network Functions (NFs), and its services. The RAN is a collection of one or more *Next Generation Node B (gNB)* responsible for establishing wireless connectivity with the UE while maintaining a physical link to the 5GC. In Fig. 1 RAN and gNB are equivalent, as the RAN consists of a single gNB.

These components exchange *Control Plane (CP)* and *User Plane (UP)* data, where the former manages the connection, and the latter handles actual user data, e.g., the industrial process data. The UE initiates connections and transmits data to the RAN which oversees various functions, including resource allocation and forwarding CP and UP data to the *Authentication and Mobility Function (AMF)* and the *User Plane Function (UPF)*, respectively. The AMF collaborates with other 5GC functions to govern the UE connection, overseeing authentication, mobility, among other functionalities. The UPF routes user data, typically to the Internet. Security controls for CP data exist between the UE and the AMF, while for UP data, they are implemented between the UE and the RAN.

5G introduces numerous novel technologies such as *Network Slicing*, which enables the deployment of multiple network instances over the same hardware with dedicated resources. Additionally, Mobile Edge Computing (MEC) enables the placement of services and resources closer to the end-user, which reduces latency. These advancements, alongside others, make 5G well-suited for industrial settings, offering benefits like enhanced reliability, lower latency, and increased bandwidth to support the increasingly demanding requirements of industrial applications [26].

*2.4. 5G in industrial networks*

Latency and reliability are critical in industrial networks, ensuring smooth and safe operations while also enabling rapid detection and mitigation of disruptions before they escalate into serious issues. Low latency facilitates real-time communication between devices, essential for timely decision-making. Reliability complements low latency by ensuring robust communication channels and redundant pathways, minimizing packet loss and increasing resilience.

This is especially vital in high-stake environments like nuclear power plant controls, where even minor communication delays could lead to catastrophic failures, potentially endangering millions. In such scenarios, systems must shut down instantly to prevent contamination or widespread harm. The capabilities of 5G to ensure *sub*-ms network latency make it a strong candidate for industrial applications. In the following, we explore the key reasons behind 5G's suitability for industry, the advantages it offers in terms of automation, optimization, flexibility, and scalability.

**Benefits of 5G over other wireless technologies.** As highlighted by Fig. 2, 5G URLLC is the sole wireless technology suitable for industrial usage. Previous mobile wireless technologies, such as LTE, provided high bandwidth but suffered from significant latency. Similarly, protocols such as Wi-Fi, despite their lower energy consumption, suffer from higher latency, making them unsuitable for industrial use [28]. Even the latest (6th) generation of Wi-Fi, which promises lower latency, falls short compared to 5G URLLC. While both technologies can achieve sub-ms performance, 5G maintains this low latency while carrying five times more data, making it more reliable for demanding industrial applications [29]. Furthermore, the same paper demonstrates that as more access points (APs) or gNBs are added to increase network capacity, Wi-Fi latency exceeds the 10 ms upper limit (cf. Fig. 2), even under low load of traffic. In contrast, 5G consistently meets low latency targets. While Time Division Multiple Access (TDMA) based protocols, such as Wireless Networks for Industrial Automation-Factory Automation (WIA-FA), can compete with 5G URLLC in terms of latency, they usually offer lower speeds, reduced coverage, and capacity [30,31]. Moreover, 5G is the first wireless network to fully support *TSN* [32], an IEEE standard ensuring predictable communication in traditional Ethernet networks through synchronization, redundant communication, and time-aware Quality of Service (QoS) [33,34]. While Wi-Fi 6 offers partial TSN support, not all vendors provide it, and standardization to fully support TSN is underway [35]. TSN's industry-wide adoption is driven by its capacity to transform less reliable Ethernet networks into deterministic, low-latency systems. Consequently, 5G stands as the only wireless technology meeting this demand. In addition, 5G offers additional significant benefits, such as *increased security* (analyzed in Section 4) and a *sub-meter* positioning system that enhances asset tracking and streamlined process optimization. Finally, 5G also offers the inherent benefits of wireless technology, such as enhanced automation and mobility as detailed in Section 2.2.

> **Summary:** 5G stands out as the only wireless technology that not only satisfies the industry's ever-growing demands but also enables innovative use cases and introduces tools that bolster functionality, automation, and cost efficiency. However, the introduction of numerous new components and technologies with 5G expands the industrial network's attack surface. Thus, ensuring the secure deployment and operation of the 5G network is paramount to safely realizing its benefits.

## 3. 5G trials in industrial environments

As discussed earlier, 5G is currently the only wireless technology fit for industrial communication (cf. Section 2.4). To further exemplify the new possibilities that 5G communication enables in the industrial sector, we discuss the real world application of 5G in the following. To this end, we discuss different use cases for industrial 5G communication [17] in Section 3.1. Then, we contextualize these scenarios by analyzing two exemplary 5G trials in a real life factory in Sections 3.2 and 3.3.

## 3.1. Use cases for industrial 5G

As a novel technology, 5G offers a wide range of applications in industrial networks, enabling the development of new use cases [17]. One significant advantage of 5G is the *mobility* it brings to previously stationary or limited-movement components, overcoming cabling constraints. UEs can now be integrated into industrial components such as sensors and robots, enabling automation and improvement of multiple tasks on the production line. If necessary, one UE can be integrated into every industrial component enabling direct wireless communication with the gNB, and thus mobility for every component; however, if mobility is not required, one UE can serve multiple industrial components, for costs saving. Examples of use cases requiring mobility include *Warehouse Automation*, where autonomous robots transport and organize goods, and *Motion Control*, which involves a closed control loop regulating moving parts in a physical process. The latter task can be particularly challenging in non-wireless scenarios [16]. Another significant aspect of 5G which enables new use cases is its low latency capabilities. Traditionally, due to the importance of real-time monitoring and control for safety, HMI systems connected by wires were positioned within the shop floor, limiting *flexibility* and increasing costs. However, with the adoption of 5G, devices on the shop floor can now transmit data in real-time to systems outside the industrial premises, enabling *Remote monitoring and Control* [26]. Moreover, *Machinery Maintenance*, which typically requires on-site specialists, benefits from 5G's real-time capabilities. Specialists can remotely guide on-site personnel through live data transmission using camera sensors, reducing costs and response time. Lastly, the innovations of 5G enable new security configurations. A major example here is network slicing, which enables effective *Network Segmentation* by dividing the network into slices with isolated traffic and distinct security configurations, thereby reducing the need for additional hardware for Virtual Local Area Networks (VLANs) such as routers and switches, currently used in industrial networks for segmentation purposes.

## 3.2. Factory-cloud based collaborative mobile vehicles

As an example of how 5G can enable self-governed mobility of vehicles in an industrial setting, researchers in the 5G-SMART project [36], in collaboration with Ericsson GmbH, developed a remote-controlled system for automated guided vehicles (AGVs) [23]. The system is built upon a 5G SA network, factory-cloud servers, and a fleet of AGVs, which can move without additional guiding lines. Each AGV scans its vicinity using sensors (e.g., a LiDAR or 3D-camera), creates a map of potential obstacles, and locates itself within the factory environment. Additionally, the AGVs forward their sensor data to the factory cloud. The cloud servers use these data to provide additional AI-based functionalities to the AGVs. This includes, e.g., a *common map*, which is the aggregate of all local maps from the individual AGVs and provides an enhanced overview of the current situation on the factory floor.

In the "standard" case of AGV systems, a vehicle can only plan its path based on local data. In contrast, this collaborative system can utilize data from the common map to plan ahead for, e.g., obstacles that cannot be seen by one vehicle, but that are already reported by another. The researches validate that the enhanced *trajectory control* and *obstacle avoidance* reduce traveling times of the vehicles by 32% on average.

In addition to these high-level commands (e.g., vehicle A move to position X), the servers can additionally provide low-level commands to the servo-motors of the vehicles. To enable this fully remote-controlled scenario, the researchers also verified the URLLC property of the 5G SA network with a median latency of $0.8\,\text{ms}$ and the 99.9th percentile of the latency being below $1.3\,\text{ms}$ [37]. Since *safety* is the most important security goal, AGVs must adhere to the safety guidelines. To this end, the researchers measured the stopping distance in an emergency scenario. With the maximum legally allowed velocity of $0.9\,\text{m s}^{-1}$ to $1.0\,\text{m s}^{-1}$ for AGVs, the system achieves a stopping distance of $13.8\,\text{cm}$, which is within the requirements [23].

Summarizing, this trial for 5G networking in industrial environments validates that URLLC is possible in an industrial scenario. Additionally, this use-case demonstrates how 5G communication enables additional possibilities for remote controlled vehicles.

## 3.3. 5G communication in a semi-conductor plant

As the second trial for 5G in a factory, the researchers of the 5G-SMART project [36] introduce 5G communication to a production line for semiconductors wafers [23].

The motivation to equip a stationary production with wireless communication is to reduce commissioning and relocation times of individual components in the production line. More flexible production line components enable operators to accommodate changing production demands more quickly and cost effectively [23]. While latency requirements are not as strict as in the previous use-case (Section 3.2), availability, the underlying network must still provide availability and timeliness guarantees to ensure optimal utilization of the production line. Therefore, a 5G SA network, that provides high throughput with eMBB, is the most suitable wireless technology for this use case.

The experimental setup of this use-case consists of a *line-controller*, which monitors several *machine-controllers*. Each *machine-controller* receives sensor data from the production line, such as, e.g., temperature and air-pressure. During operation, the *line-controller* periodically polls each *machine-controller* for its current state. In their tests, the researchers compare the performance of this system with the 5G SA network with an Ethernet based setup. The results show that Ethernet communication outperforms 5G in terms of reliability, timeliness, and overall network quality. However, the researchers also report a substantial increase in flexibility for the production line. Specifically, they show a decrease in relocation time from $12\,\text{h}$ to $0.5\,\text{h}$ and a reduction of commission time from $8\,\text{h}$ to $1\,\text{h}$. Furthermore, they also calculate annual cost savings[1] of estimated €237.500 for this production line, when switching from Ethernet to 5G [23].

Summarizing, this trial for 5G shows that, although 5G communication might not provide the same performance guarantees as, e.g., Ethernet-based communication, there are considerable benefits for flexibility and cost reduction.

> **Summary:** 5G communication provides a wide range of different use-cases in the industrial sector, such as, e.g., *mobility* of components and *flexibility* for production lines. Real-World trials demonstrate the benefits and new possibilities that 5G enables. Specifically, validating sub $1\,\text{ms}$ transmission for URLLC.

## 4. State of the art: Secure usage of 5G

The 5G specifications enable the deployment of *non-public 5G* networks (also known as *Private 5G*) owned and managed by individual companies, allowing them to leverage the advantages of 5G without impacting commercial networks. However, the complexity inherent in the 5G specifications poses challenges for secure integration. In addition, specifications are often complicated and ambiguous, dispersed across multiple sources, and many of its controls are optional to utilize and implement. To facilitate an easier and secure deployment of 5G networks in industrial environments, we present the state-of-the-art in 5G secure integration, including deployment options that focus on the physical setup of the network (Section 4.1). Furthermore, we discuss configuration techniques, which involve the customization of network settings in terms of security (Section 4.2), thereby focusing on the specifics of using 5G in industrial networks.

---

[1] This estimation does not include cost for the initial investment for the 5G infrastructure. Instead, it compares the running costs of an Ethernet-based production line with a 5G production line.

*4.1. Secure deployment*

The deployment of a 5G system is a crucial step on securely integrating 5G in the industrial network. This section explores different deployment options for private 5G networks, including SA versus NSA configurations (Section 4.1.1), as well as various private 5G topologies (Section 4.1.2). Choosing the right security setup can be challenging for industrial companies new to cellular networking. As a result, cost often becomes the main focus, especially when the risks of using cheaper NSA deployments are not fully understood. The section also delves into the security considerations associated with these deployment options.

*4.1.1. Standalone and non-standalone*

3GPP explicitly defines two deployment options for a 5G network: *standalone (SA)* and *non-standalone (NSA)* deployments. The SA deployment, as detailed in Section 2.3, is considered *"the true"* next generation of mobile networks. It incorporates all the innovations and improvements outlined in the 5G specification. On the other hand, the NSA deployment utilizes a 4G core network, also known as Evolved Packet Core (EPC), instead of a 5GC, and a RAN consists of one or more evolved Node B (eNB; the 4G equivalent of gNB) and one or more gNB [9]. The UE is connected simultaneously to both a gNB and an eNB, both of which are connected to the 4G core. Here, the eNB is responsible for handling the CP data of the connection, while the gNB handles the UP data. The purpose of this deployment was to enable a smooth rollout of 5G, allowing end-users to already use the high bandwidth offered by 5G in early stages. In this paper, when we refer to 5G, we imply a 5G SA deployment.

Regarding security, SA deployments incorporate all the security enhancements provided by 5G. In contrast, NSA relies on the 4G security specifications, which are notably inferior to those of 5G [9]. The most significant security improvements of SA over NSA are outlined in Table 1. To begin with, in a 5G SA network, all subscriber credentials are encrypted, and only the 5GC can decrypt them. This setup effectively prevents attacks related to user privacy, such as tracking their location (more information in Section 4.2.4). Moreover, integrity protection and encryption are mandatory features to support for the UP, as opposed to being optional as in the case of 5G NSA. In industrial settings, where the UP carries all industrial traffic, these features are crucial for preventing attacks like data tampering and false data injection [38]. Another crucial feature of 5G SA is the encryption of the initial Non-Access Stratum (NAS) message, which is the protocol responsible for CP signaling between the UE and the AMF. This serves as a defensive mechanism against DoS attacks towards the UE [39]. Furthermore, in a 5G NSA network, the UE radio capabilities are transmitted without protection between the UE and the RAN. As the name suggests, this message contains critical information about the UE's radio capabilities, such as supported frequencies. Tampering with this information can lead to undesirable outcomes, notably battery drain, which can be catastrophic for ICSs with battery-powered sensors [9]. 5G SA counters this by sending this information after the CP security establishment, which means that at least integrity protection will be applied (CP security is discussed in Section 4.2.1). On top of those benefits, 5G SA provides support for 256-bit cryptographic algorithms for both UP and CP for enhanced security and protection against potential quantum-attackers [40].

However, security is not the only concern with NSA deployments in an industrial context. NSA fails to meet industry demands for low latency and real-time operation, particularly in the context of URLLC, as it relies on an EPC that was not designed with low latency in mind. Additionally, it lacks essential functionalities for industrial networks, such as TSN support. Even-though 5G NSA could be utilized by industrial companies already possessing a 4G network and intending to transition gradually switch to a 5G network, companies should not rely on it due to its weaker security and performance. For instance, the system for collaborative mobile vehicles (cf. Section 3.2) URLLC is necessary to meet the safety requirements. To summarize, 5G SA is the true 5th generation of mobile networks, able to meet the demands for low latency and high security.

**Table 1**
NSA and SA security.

| Security control | 5G NSA | 5G SA |
| --- | --- | --- |
| Subscriber Identifiers | No protection | Send encrypted |
| UP Security | Option. support | Mandat. support |
| Initial NAS Message | No protection | Send encrypted |
| UE Radio Capabilities | No protection | Send protected |
| Crypto. Algorithms | 128-bits | 256-bits support |

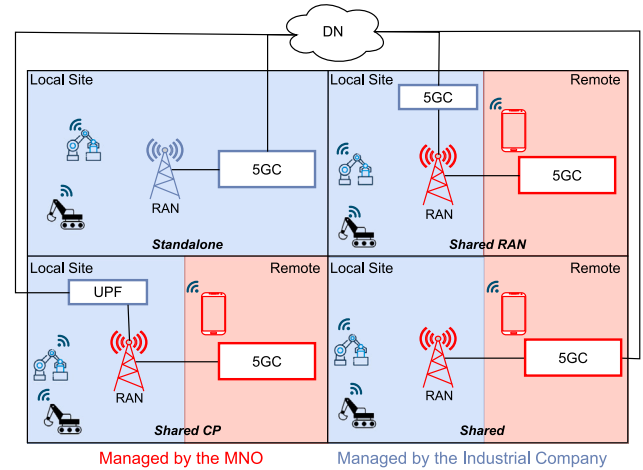

**Fig. 3. The different Private 5G deployment models:** Besides standalone deployment, where the industrial company owns and manages the entire network, other deployment scenarios involve sharing parts with MNOs, including elements such as the RAN and parts of/the entire 5GC. This sharing is facilitated by Network Slicing.

*4.1.2. Private 5G topology options*

The 5G specification [25] defines two types of 5G networks: public 5G networks operated by Mobile Network Operators (MNO) and non-public (or private) 5G networks managed by private organizations or corporations. While public 5G networks are deployed to serve commercial customers and are typically configured to meet the requirements of average end-users for eMBB, private networks can be tailored to address the specific needs of industries, usually focusing on the URLLC and mMTC. Various deployment models exist for private 5G networks, each offering distinct advantages and benefits over the others. Deployment methods include choosing between licensed, unlicensed, or shared spectrum, as well as incorporating ownership and management of the components between the enterprise and a MNO [11]. In the following, we present the four deployment options for private 5G deployments identified by 5G-ACIA [10], as depicted in Fig. 3. Additionally, we qualitatively compare these four deployments in terms of management, cost, performance, and security, and present our results in Fig. 4.

As illustrated in Fig. 3, the *Standalone* (distinct from 5G SA) deployment is a scenario in which the industrial company deploys and controls all components of its private 5G network on its own. In contrast, the *Shared RAN* deployment involves sharing the RAN with the MNO. The RAN is connected to both the MNO's and the industrial company's 5GCs, but it is usually managed by the MNO. Similarly, in the *Shared CP* deployment, the two entities additionally share the CP of the 5GC. Lastly, in the *Shared* deployment, the industrial company utilizes the MNO's existing infrastructure, where the MNO handles both RAN and 5GC, achieved through an agreement for a dedicated network slice [41].

Each of these deployments offers unique benefits, but only a standalone deployment holds the potential for increased security levels and high performance. With full control over every component, an industrial company can enforce comprehensive security controls, such as encryption protocols, segmentation techniques, and access control, on every 5G component. Additionally, the dedicated resources ensure that the entirety of the resources will be available in case of need [42]. The Shared RAN deployment, involving a RAN sliced in two or more slices, could theoretically provide similar performance and security. However, by sharing components such as the RAN, the risk of unauthorized access to industrial networks is increased. As UP security controls, are only between the UE and RAN, the encryption keys are stored within the RAN. A malicious MNO that manages the RAN and has physical access to it, could potentially gain access to UP encryption keys [43]. Furthermore, successful attacks on one slice with inferior security controls, could potentially affect the other slices, e.g., by draining all the available resources, if strong isolation controls are not in place. In context of the real-life 5G trials (Section 3), a compromised shared RAN could lead to substantial information leaks. For example, in-depth knowledge of vehicle and worker behavior in the collaborative vehicle use-case.

The remaining two deployments are unsuited for industrial usage as they face performance issues, i.e., they do not meet industrial requirements for low latency and real-time operation. Specifically, the *Shared* deployment encounters performance challenges due to the remote placement of the 5GC (including the UPF) and is thus not suitable for industrial usage as it might not meet low latency requirements. On the other hand, the Shared CP deployment's remote placement of the 5GC CP should not be an issue for time-critical applications. However, in certain cases with a significant amount of CP traffic, such as in mMTC scenarios, it may still become problematic [11]. For example, congestion in the 5GC can occur due to multiple authentication requests, such as in an IIoT scenario with multiple devices. As the 5GC CP authenticates each device before allowing access to a specific slice and its dedicated resources, authentication requests from multiple slices[2] are handled by the same underlying resources. This could throttle resources and introduce significant latency in the authentication of industrial devices, as authentication is one of the most expensive operations in 5G [44]. In addition, both of these deployments are significantly less secure than the other deployments: Having a remote 5GC means that sensitive data, such as authentication and/or UP data, is being sent, processed, and stored outside the company's premises. This increases the risk for unauthorized access, tampering, and other security concerns [45].

While a complete local deployment of the 5G network is the preferred option for enhancing security in industrial settings, it is essential to indicate that the effectiveness of the Standalone deployment heavily relies on *"the optimal"* (regarding security) configuration of network settings. In the absence of such configuration, the security potentials of a Standalone deployment are compromised.

### 4.2. Secure configuration

5G offers a range of security enhancements compared to previous mobile network generations, making it increasingly appealing for use in industrial networks. However, many of the additional security controls are optional. In the following, we filter out and explain optional controls deemed important for industrial networks [13,14].

---

[2] This includes the industrial company slice, the MNO slice, and other slices that the MNO may host.

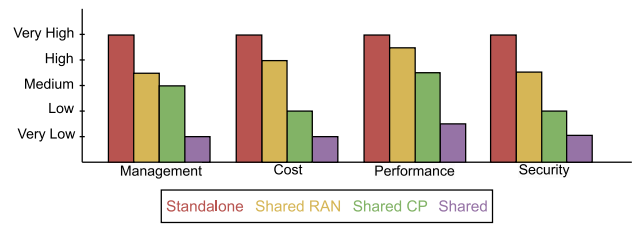

**Fig. 4. Capabilities of the different private 5G deployments:** The capabilities of each deployment vary in terms of management, cost, performance, and security. Generally, the more expensive the deployment, the easier it is to manage and ensure high performance and security.

#### 4.2.1. Control plane and user plane data security

In a 5G system, data is split between two planes: Control Plane and User Plane (cf. Section 2.3). CP data is exchanged between network components to manage various aspects of communication, e.g., call setup, handovers, network registration, and resource allocation. UP is the actual data that users send.

The 5G specification outlines a set of four pairs of cryptographic schemes for encryption and integrity protection for UP and CP data [46]. These schemes are based on different cryptographic algorithms: *NULL*, *SNOW*, *AES*, and *ZUC*. It is important to note that the NULL scheme offers no protection at all, and if used, data is transmitted as plaintext and without integrity protection. In addition, according to the specification, implementing these schemes is mandatory for both planes, expect the pair based on ZUC, which is optional. While support for integrity protection and encryption is mandatory for both planes, their utilization is optional and at the discretion of the operator [13]. In other words, the operator may choose to utilize the NULL scheme. However, an exception lies in the mandatory integrity protection for CP data, where the NULL scheme is not permitted (with certain exceptions such as emergency calls). In industrial settings, prioritizing the use of optional security controls for both planes — avoiding the NULL scheme — is crucial due to various concerns associated with each plane.

To begin with, it is noteworthy that industrial protocols rarely operate over security protocols such as TLS or IPSec [47]. Consequently, industrial data often remain unprotected over the underlying network, in our case, the UP of 5G. Even if a security protocol is used, it is usually one of the previously mentioned protocols on the upper layers of the protocol stack, which do not offer any protection to TSN data, as TSN operates at the data link layer. Thus, protection needs to be applied either at the data link layer or the physical layer. Therefore, as TSN data are also being encapsulated in the UP of 5G, tampering attacks on the synchronization data may remain feasible and threatening with a complete loss of availability, even if TLS or IPSec is used. To address this, optional 5G security controls should be enforced to provide protection to the UP data. In an industrial setting, the integrity of the UP data is deemed as the most important security control, as without it, an attacker could intercept, inject, and manipulate industrial data (such as controller commands and sensor readings) or synchronization data, causing a system malfunction or a complete loss of availability, and potentially threatening human lives. While encryption is also important, it is less critical for industrial networks. Encryption of the UP safeguards against exposing the system's information to unauthorized parties but does not directly protect the safety, e.g., of the personnel. Similarly, encryption of the CP will safeguard the data from unauthorized parties, ensuring that an attacker is not able to monitor the network and extract information about the 5G network and identify flows/vulnerabilities (such as the AMF accepting authentication requests with the NULL scheme for integrity protection of the CP [48]).

In 5G networks, the presence of the wireless link (i.e., the Uu interface) between the UE and RAN further emphasizes the need for utilizing encryption and integrity protection controls, as radio waves are much easier to intercept than bits on the wire. Anyone with cheap radio equipment (also known as sniffers) could potentially eavesdrop on UP/CP data [49]. While it could be argued that manipulating over-the-air data is much harder than eavesdropping and requires extensive knowledge and tools, it is still possible. For example Rupprecht et al. successfully manipulated UP data in 4G networks where UP integrity protection does not exist, by utilizing Software-Defined Radios (SDRs) and open-source software [50]. Consequently, integrity protection and encryption are important for both CP and UP, and should thus be enabled.

*4.2.2. RAN internal interfaces & N2/N3 security*

Besides the wireless interface between UE and RAN, multiple interfaces in the 5G architecture require support for security mechanisms, yet their usage is optional. Regarding the RAN internal interfaces (green section in Fig. 1), all interfaces should support the IPSec ESP protocol for confidentiality, integrity, and replay protection, along with IKEv2 for authentication [46]. However, the specification mandates their usage only for the F1-C and E1 interfaces, handling sensitive control and management data. The application of these protocols to the F1-U interface, responsible for transmitting UP data, remains optional [13].

Similarly, the N2 and N3 interfaces, responsible for transmitting CP and UP between the RAN and the 5GC respectively, must support IPSec with IKEv2, and in the case of N2 also Datagram Transport Layer Security (DTLS). However, the decision to actually configure the 5G system to utilize them is left to the operator once more [13,46]. While CP data benefits from mandatory integrity protection between the UE and the AMF, safeguarding it over the N2 interface, UP data do not benefit from any security controls of the UP. Even with the optional UP integrity protection and encryption in place, UP data remain exposed over the N3, as security controls for UP only protect the data between the UE and RAN over the Uu interface. Hence, leveraging all available optional security protocols is crucial to ensure encryption and integrity protection to the UP traffic over N3. Since integrity protection safeguards against, e.g., command injection, omitting these optional security measures may lead to considerable damage in real world scenarios. In context of the two presented 5G trials (cf. Section 3), e.g., malicious commands could tell the production line to discard a semiconductor wafer that is actually fine, causing monetary loss. Alternatively, in context of the collaborative vehicle system, false commands could potentially even cause crashes and thus destroy equipment or even harm personnel.

*4.2.3. Core network security*

After examining the RAN internal interfaces and the interfaces connecting it to the 5GC, we proceed to examine the 5GC itself. For the core network, 5G introduces a Service-based Architecture (SBA) comprising interconnected Network Function (NF) that cooperate with each other to handle various UP and CP functions. These functions expose their services, through RESTful APIs.

A key benefit of this architecture is its modularity and ability to integrate customized, on-demand additional NFs to the 5GC. These NFs can potentially be exposed to third parties to expand the services of the 5GC [51]. However, this exposure increases the risk of malicious or compromised NFs substantially. In such a case, the malicious NF could potentially disrupt operation of the 5G system or exfiltrate sensitive data. Hence, prioritizing security controls like authentication, authorization, and end-to-end security across SBA interfaces is crucial.

Authentication prevents the deployment of unauthenticated NFs by verifying their identity, while authorization ensures that only authorized NFs can access certain services and data based on predefined permissions. End-to-end security provides protection against data tampering and eavesdropping. To meet these security requirements, 3GPP mandates the implementation of TLS for authentication and end-to-end security, coupled with OAuth 2.0 for authorization [46]. While the use of TLS (or alternative controls [13]) for establishing encryption, integrity protection CP data, and entity authentication over the SBA interfaces is mandatory, the use of OAuth is not. The OAuth authorization framework leverages tokens to grant different access levels between consumer NFs and service NF providers, thereby preventing unauthorized access to critical functions, such as retrieving user identifiers or setting network configurations and policies. Unauthorized access to NF services can have numerous consequences, ranging from extracting sensitive data to a complete loss of availability for the ICS, e.g., by deregistering other NFs such as the AMF from the 5GC [52].

Consequently, to ensure the security of the 5GC, both TLS and OAuth 2.0 should be used. Any alternative to TLS should be well-justified, offer at least the same security controls, and still adhere to latest security standards.

*4.2.4. Other optional security controls*

Assuming that an industrial company employs all optional controls outlined in the preceding sections, we could contend that both UP and CP are safeguarded across every interface, rendering the architecture as secure as possible. Nevertheless, there are still parts of the communication that remain unprotected, and other additional non-compulsory security controls exist that can further enhance the security of the industrial network if utilized.

**SUPI Encryption.** As briefly mentioned in Section 4.1.1, the subscriber's credentials in a 5G network are sent encrypted during authentication, preventing localization, linkability, and tracking attacks. These credentials include the Subscriber Permanent Identifier (SUPI), which is the unique identifier assigned to each UE for identification. In previous mobile network generations, such as LTE (where it is known as IMSI), IMSI catchers posed a significant threat by allowing the tracking of individuals since the identifier was transmitted in plain text [53,54]. To mitigate this risk in 5G, SUPI encryption was implemented, but only as an optional control.

In the context of industrial 5G networks, a linkability attack occurs when an attacker is able to correlate different sessions or activities back to the same UE. This allows the attacker to identify patterns in industrial communication and gain crucial information from traffic (data or metadata in the case of encrypted UP data). For example, by analyzing the traffic, an attacker may deduce which UE corresponds to a controller or sensor and attempt to interfere with the connection. Although linkability attacks have been proven possible even with SUPI encryption, these attacks do not pose as significant a threat as IMSI catchers [55].

Similarly, tracking attacks involve monitoring and following the movements and activities of an industrial machine or person over time. In industrial 5G networks, an attacker may try to ascertain the presence and location of specific employees or critical industrial devices in a factory, risking unauthorized access to data or processes. For example, tracking the location of security personnel or autonomous, self-driving vehicles in a factory (cf. Section 3.2) could enable an attacker to illegally enter the premises or tamper with industrial robots. SUPI encryption helps to protect against tracking attacks by ensuring that the SUPI is not transmitted in plaintext, making it more difficult for attackers to track UE movements.

**Authentication.** 5G supports multiple authentication levels: *(a) Primary authentication* for mutual authentication between the UE towards the 5GC [46, Clause 6.1], *(b) Secondary authentication* to authenticate the UE to external networks [46, Clause 11.1], and *(c) Network slice-specific authentication* for authentication between a UE and a network slice [46, Clause 16.3]. Primary authentication can use 5G-AKA, EAP-AKA, or other EAP-based authentication schemes like EAP-TLS [56], and is the only mandatory authentication procedure. Secondary and slice-specific authentication utilize the EAP framework defined in RFC 3748 [57] and are both optional.

The primary authentication takes place during the initial access of the UE to the 5GC system, and can be invoked periodically for re-authentication of the UE. As this procedure is mandatory, the 5G system ensures that only authenticated devices can access the network. However, employing the secondary and the slice-specific authentication will further enhance the security of the system, especially in certain use cases in industrial networks. As mentioned before, secondary authentication can be employed to authenticate the UE to external networks. After the primary authentication, secondary authentication can be invoked with another set of credentials different from the primary authentication, such as digital certificates or usernames and passwords, depending on the EAP-based authentication method. This can be extremely useful for an industrial company to relegate access to external resources, such as cloud-based services or data stored in the intranet. Similarly, slice-specific authentication can be used after primary authentication to authenticate the UE to a specific network slice with a different set of credentials. This authentication is particularly important when network slicing is used in industrial networks, such as for segmentation purposes. This ensures that only authorized UEs can access resources dedicated to a specific slice, thereby mitigating risks such as unauthorized access to services belonging to a specific slice, resource draining and performance degradation.

**Table 2**
5G opportunities for enhancing security and future research directions.

| Opportunities | Research directions |
| --- | --- |
| Effortless segmentation | • Study and Enhance the scalability of a sliced-based network segmentation<br>• Develop secure low latency inter-slice communication schemes |
| Reduced downtime and enhanced availability | • Develop automated resource management techniques and dynamic scheduling (e.g., in the form of xAPPs)<br>• Establish resilient virtualization architectures<br>• Reduce latency associated with virtualization<br>• Develop agile re-deployment techniques (e.g., VM migration) |
| Distributed security at the edge | • Design multilayered, dynamic security policy enforcement mechanisms<br>• Optimizing IDSs and IPSs to utilize O-RAN capabilities |

**Further Optional Controls**. In the previous sections, we analyzed the most critical optional security controls for the industrial networks. Other optional security controls exist, such as gNb certificates and enrollment or implementation of 256-bit algorithms, which can be utilized to further enhance security, beyond from what we already proposed. More information on these controls can be found in [13,14].

> **Summary:** 5G offers many enhancements in security compared to previous generations of mobile networks, which can be utilized to enhance and retrofit security in industrial networks. Such enhancements include multi-authentication schemes and protection of UP data. However, 5G's complicated specifications, with numerous optional controls and various deployments of private 5G networks, may challenge industrial companies with no previous experience or expertise in mobile networks to deploy 5G securely.

## 5. Opportunities to enhance security

Building on the foundational security improvements that 5G provides, we believe its novel approaches and tools hold significant potential to further enhance the security of industrial networks. In this section, we explore how 5G's innovative features and modular design can be leveraged to boost security, including on current trends like Zero Trust and Artificial Intelligence (AI) integration. Capabilities such as enhanced network segmentation, real-time attack detection, and prevention at the network edge could be realized with 5G, offering promising opportunities to strengthen the overall security of industrial networks (cf. Table 2).

### 5.1. Effortless network segmentation

Segmentation in industrial networks is a commonly employed technique to divide the network into smaller segments [1]. This allows for better control of flows between different segments and helps mitigate the risks of attacks on one segment spreading throughout the network. Traditional segmentation techniques include VLANs, firewalls, as well as physical separation.

5G network slicing (cf. Section 2.4) has the capability to replace or work in parallel with traditional network segmentation techniques, enhancing security and automation in industrial networks simultaneously. Achieving a sliced 5G network involves segmenting the RAN and 5GC for the deployment of end-to-end isolated slices over the same hardware. Each slice can then have unique requirements in terms of resources, security controls, and QoS policies.

Industrial companies often separate their IT network — office network connected internet and is used employees — from their OT network, which manages the ICS systems that have strict performance and security requirements. This separation is achieved through dedicated equipment such as firewalls and routers. This set of separation devices is commonly referred to as the Demilitarized Zone (DMZ). However, with 5G network slicing, this separation can be achieved effortlessly by deploying two distinct slices — each configured differently according to the specific needs — over the same physical infrastructure, without the need for additional equipment.

However, the true advantage of 5G slicing lies in its potential to enable novel segmentation techniques, such as nano-segmentation [58]. Nano-segmentation is achieved by allowing every node in the network to verify that each packet processed *(a)* belongs to a whitelisted flow and *(b)* originates from an authorized host. This approach facilitates per-device segmentation, establishing a Zero Trust environment where no device is trusted by default, and every industrial UE is isolated within its own slice with dedicated resources.

In a 5G industrial system, we could potentially achieve nano-segmentation by leveraging slicing-based segmentation and refining the approach through reducing the size of each network slice, potentially down to a single UE per slice. This approach is depicted in Fig. 5 where the industrial network, consisting of three components, is divided into three slices, one for each component. By doing so, each slice becomes a highly isolated environment, ensuring that each UE operates within its own dedicated and secure virtual network. This granular segmentation significantly mitigates the risk of lateral movement by malicious actors within the network. Slice-specific authentication mechanisms can be utilized to verify the identity of the UE within each slice. In this way, each network node can confirm that packets were generated by an authorized UE by checking the source slice of the packet, as only one UE has access to the corresponding slice. Furthermore, NFs can be implemented at various levels of the network, including the RAN, 5GC, MEC nodes, and Software Defined Networking (SDN) Controllers (cf. Section 6.2), to enforce allowlisting policies. These policies ensure that only authorized communications are permitted between slices (and therefore between industrial devices), thereby maintaining compliance with the overall network security framework.

Additionally, a slice-based segmented network could further enhance automation and security. This approach would allow each slice to allocate or deallocate resources as needed. In the event of an attack, the 5G system could potentially denylist specific slices, quarantining affected devices before the threat spreads within the network [59].

However, more research is needed, as network slicing-based segmentation may be constrained by the network's scalability to manage a growing number of slices efficiently. As the number of slices increases, ensuring optimal performance and resource allocation becomes more challenging. Additionally, direct communication between devices in different slices is not currently permitted, and traffic must be routed through the core network. For low-latency communications, research should focus on establishing secure cross-slice communication channels that bypass the core network. This might involve application-level communication via cloud-based servers located within the RAN, enhancing both the efficiency in cross-slice interactions.

*5.2. Reduced downtime and enhanced availability*

Safety and thus availability are the most important security requirement of an ICS, as any interruption may cause widespread disruptions and potentially threaten human lives. For example, in the case of power grid systems, an availability issue could result in large-scale power outages, affecting hospitals, transportation systems, and emergency services [60]. Similarly, in industrial manufacturing, an interruption could halt production lines, leading to significant economic losses and potential safety hazards for workers. The virtualized architecture of 5G could be utilized to enhance the availability of an ICS and mitigate the downtime of the system in the event of attacks or security updates.

Virtualization is a fundamental process in 5G, involving the software-based transformation of a system. Two key examples of virtualization are *Network Function Virtualization (NFV)* and *SDN*, which play crucial roles as enablers of network slicing [61]. NFV involves the virtualization of specific network functions, such as the 5GC or the RAN. On the other hand, SDN focuses on virtualizing the network management, including routing functionalities, detached from the underlying hardware. These software functions, are also known as Virtual Network Functions (VNFs). Virtualization opens the door to numerous security enhancements that could potentially reduce the downtime of the system in case of attacks or updates.

Firstly, its programmable nature enables dynamic adjustments of security mechanisms, such as logging, authentication, and verification [62]. In addition, update and patch management are simplified, as new patched versions of specific functions can be deployed in parallel with existing ones. This is crucial for ICSs, as it cannot afford frequent disruptions to its operations for security patches and updates. This is one of the major reasons why industrial equipment is often not up-to-date. Secondly, virtualization could potentially become a major security response to control attacks. Techniques such as Virtual Machine (VM) migration could be employed in scenarios where a part of the system is compromised, or a specific segment of the underlying network is under a DoS attack, to mitigate the effects and reduce or potentially avoid downtime [63]. Furthermore, as specialized hardware is no longer required, it is easier to maintain real-time updated, standby backup copies, which can be easily deployed to take over crucial functionalities, in case of a successful attack. Lastly, virtualization, as a well-studied technology, benefits from a multitude of existing research which examines techniques to mitigate and prevent DoS attacks [63–65] and to improve intrusion detection [66,67].

While virtualization has the potential to enhance security, more research is needed under the 5G concept and within ICSs. This includes exploring enhanced VNFs tailored for ICS applications, investigating automated resource management techniques, ensuring interoperability between virtualized and legacy systems, and designing resilient virtualization architectures capable of withstanding attacks. Furthermore, research should focus on reducing latency associated with virtualization, developing dynamic security policies that adapt to operational changes, and integrating MEC to deploy security or time critical services closer to the data source.

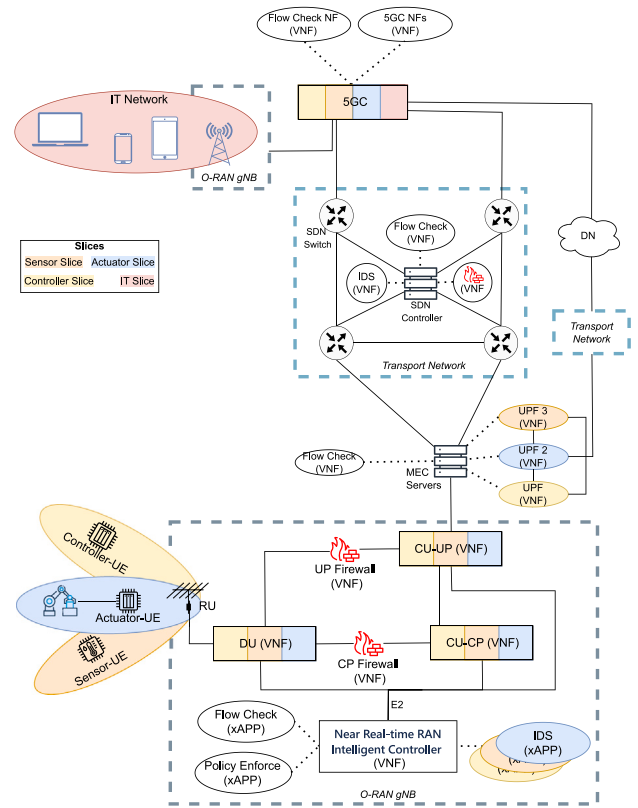

**Fig. 5. Exemplary Nano-segmented, sliced 5G network with distributed security controls:** The 5G infrastructure supports four different slices, with colors indicating slice ownership. An industrial company's network has a dedicated slice, while nano-segmentation is applied within the industrial network, allowing each industrial UE to operate in its own slice. Flow validation occurs at each 5G network node—the RAN via xAPP, and the 5GC, MEC, and SDN-controller via a custom NF. Security controls, such as firewalls and IDSs, are distributed as VNFs on shared hardware or as dedicated xAPPs.

*5.3. Distributed security at the edge*

One important innovation of 5G is the promotion of Mobile Edge Computing (MEC). This approach to the distribution of resources supports relocating computation units and data storage closer to end-users (e.g., Section 3.2), serving as a key component for enabling URLLC deployments in 5G. An example is the placement of the UPF closer to the end-user, either on MEC servers close to the RAN or even within the RAN.

In-RAN deployments of NFs could benefit from the O-RAN architecture, an extension of the 5G RAN architecture. The 5G RAN includes components such as the Central Unit-User Plane (CU-UP) and Central Unit-Control Plane (CU-CP), which handle UP and CP data, respectively, and the Distributed Unit (DU), which manages radio resources and executes lower-layer protocols. O-RAN introduces additional components like the near-real-time RAN Intelligent Controllers (RICs), allowing for dynamic deployment of AI applications (*xAPPs*) directly within the RAN. These xAPPs enhance network performance, security, and flexibility by providing real-time analytics and control capabilities over the E2 interface [68].

These AI-driven applications monitor various aspects of RAN performance, such as traffic patterns, latency, and throughput. Beyond monitoring, these xAPPs, powered by advanced Machine Learning (ML) models trained and hosted within the RICs, can apply real-time control mechanisms to optimize RAN functionality. This may include dynamic resource allocation and scheduling, prioritization of certain traffic types, load balancing, or even dropping connections that are deemed non-critical or potentially malicious, functions crucial for maintaining high network performance and security in real-time.

Combining MEC with the modular architecture of 5G, which splits the 5GC and RAN into multiple components, an industrial company could establish a distributed security scheme. This scheme would enforce individually tailored security controls on each network component at multiple levels, particularly at the edge of the network.

This could enable on-time detection and response to attacks before affecting critical components and the spread of lateral movement. For example, within the RAN, different firewalls could be deployed on the CU-UP and CU-CP — components of the RAN that handle UP and CP data, respectively — to establish distinct security policies for the two planes.

Additionally, *Intrusion Detection System (IDS)* and *Firewalls* could be implemented as xAPPs in the RAN to detect attacks in near-real-time [69–71]. IDSs and Firewalls are commonplace in industrial settings due to their ability to retrofit security in insecure systems [1]. They take advantage of the deterministic traffic patterns typical in ICS environments, enabling them to effectively identify anomalies and potential threats [72]. The 5G O-RAN architecture facilitates easy expansion of these systems. For example, upon detection of an attack, an intrusion detection xAPP could send commands to the RAN components (CU and DU) in near-real-time to drop the connection, thus mitigating the impact of the attack.

Furthermore, *Honeypots* [73], systems that simulate the ICS to attract attacks towards themselves instead of the real ICS, could be deployed alongside the UPF in the MEC servers. This not only serves to divert and analyze potential threats but also provides an additional layer of protection for the legitimate ICS.

In conclusion, integrating MEC within the 5G O-RAN architecture offers a unique opportunity to implement AI-driven security at the network edge. To fully capitalize on this potential, several research directions are recommended: developing multilayered, dynamic security policy enforcement mechanisms; optimizing IDSs and Intrusion Prevention Systems (IPSs) using advanced machine learning techniques that leverage the O-RAN's E2 interface for real-time analytics and node control; and studying the impact of MEC on both performance and security.

> **Summary:** 5G has the potential to further improve the security of ICS. Its novel technologies and approaches can complement and enhance traditional security controls. Network segmentation becomes effortless with network slicing, and critical functions such as IDS can be deployed instantly anywhere in the network. Distributing these security controls throughout the ICS, appropriately placed and configured for each component, enables threat prevention and detection at the network's edge, enhancing availability.

# 6. Challenges and research directions

Despite the opportunities 5G provides to enhance security, 5G also brings new challenges for ICS security. The enormous complexity and new technologies introduced by 5G significantly expand the attack surface of ICSs. Many concerns are present, particularly regarding two critical aspects for the industry: availability and real-time operation. In the following, we detail these concerns, highlighting the need for further research to address them and pointing out potential directions for future work. We summarize our findings in Table 3.

## 6.1. Jamming attacks on the wireless interface

While 5G offers significant security advantages, it is not immune to physical layer attacks. *Jamming*, where attackers disrupt communication by causing interference in the wireless channel, remains a threat. Although 5G promises improved resilience compared to LTE [74], a sufficiently powerful jammer can still cause significant disruption. For industrial networks, jamming poses a critical risk. A successful attack could completely sever communication between UE and RAN, effectively rendering the system unavailable [75]. Since complete prevention is not feasible, industrial operators must focus on detection and mitigation. Fortunately, 5G offers built-in mitigation techniques. Redundancy communication channels for UP data can be established through additional gNBs and UPFs operating on different frequencies.

However, research in this area continues. Different works, such as Arjoune and Faruque [76] or Barros et al. [77], propose further countermeasures such as frequency hopping and dynamic scheduling to strengthen 5G's resilience against jamming attacks. Frequency hopping involves changing the carrier frequency of the transmitted signal according to a pseudo-random sequence, making it difficult for jammers to target a specific frequency. Dynamic scheduling adjusts the allocation of resources according to the current network conditions. By doing so, the system can avoid predictable patterns of transmissions, that jammers might exploit. In addition, other countermeasures such as Direct Sequence Spread Spectrum (DSSS) [78], which spreads the information signal over a bandwidth larger than required, hold the potential to provide significant protection against jamming attacks. DSSS spreads the signal by multiplying it with a pseudo-random noise sequence, making the transmitted signal to look as noise to unauthorized receivers and reducing the impact of narrowband jamming. While DSSS might be unsuitable for commercial networks that aim to maximize bandwidth capacity, it could be applied to industrial networks that usually do not require as much bandwidth.

Complementing approaches to what is already proposed, could be the integration of an IDS tailored for wireless communications. As mentioned in Section 5.3, IDSs are often deployed in industrial networks to retrofit security, by alerting the operators about potential ongoing attacks. As these IDS often take into consideration deterministic characteristics of the industrial traffic, such as the inter-arrival time of packets containing industrial data [79], they could also potentially identify jamming attacks. Jamming attacks will impact industrial traffic, for example, by increasing the drop rate, latency, and inter-arrival times. In addition, an IDS system in the form of an xAPP could also utilize RAN or UE reports [71], such as the Signal to Noise Ratio (SNR) metrics produced by the gNB, to identify potential jammers in the area. Jammers can cause significant degradation in the SNR of a 5G cell due to radio wave interference. However, further research is needed to determine whether traditional IDS methods, such as the aforementioned inter-arrival-time IDS, work effectively in wireless environments and detect jamming. Additionally, research should explore how modern approaches, such as RAN xAPPs, can enhance the detection of jamming attacks.

## 6.2. Virtualization impacting availability

Virtualization is a core element of 5G networks, involving the virtualization of both NFs and the network infrastructure itself. However, by transitioning to a software-based network, the points at which an attacker can target the system increase exponentially. Software bugs, runtime errors, and additional single points of failure are introduced into the industrial network, making it harder to secure, especially regarding availability. Therefore, despite its potential, as discussed in Section 5.2, securing the 5G industrial network and ensuring its high availability requires significantly more effort.

**Virtualization of NFs:** The general threats posed by virtualization also apply to the ICS. The digitalization of network functions introduces risks associated with software vulnerabilities, which can be difficult to address. Issues include software bugs, known vulnerabilities in open-source tools, runtime vulnerabilities, and insufficient input sanitation. These threats could enable attackers to exploit weaknesses and gain unauthorized access to the 5G system [62,80]. To address these issues,

**Table 3**
Challenges of 5G affecting real-time operation and availability of ICS, and future research directions.

| Challenges | Availability | Real-time operation | Research directions |
|---|---|---|---|
| Jamming attacks | ✓ | ✓ | • Strengthening PHY layer via techniques such as Frequency Hopping, DSSS, and Dynamic Scheduling |
| Virtualization of NFs | ✓ | ✓ | • Development and standardization of secure programming frameworks, guidelines and robust isolation controls |
| Cloud deployments | ✓ | ✓ | • Development of robust cloud-specific DoS mitigation techniques |
| Centralized control | ✓ | – | • Research on distributed, and low-latency architectures as alternatives to centralized systems |
| Fake base stations | ✓ | – | • Development and standardization of low latency authentication schemes |
| Performance over security | – | ✓ | • Evaluation and standardization of modern and faster cryptographic algorithms |

*frameworks need to be developed* that enable secure programming practices for virtualized NFs. Isolation mechanisms should be standardized at every layer to prevent resource draining or privilege escalation by compromised NFs or slices [81,82]. This is critical for avoiding severe consequences, such as, e.g., the complete loss of availability in industrial network deployments.

**Cloud Deployments:** The virtualization of 5G components makes cloud deployment of NFs attractive for many companies due to potential cost savings. However, hosting critical NFs in cloud services such as Amazon AWS or Microsoft Azure exposes them to Distributed Denial of Service (DDoS) attacks, which can lead to a total loss of availability [83], jeopardizing the operation of the ICS. For cloud deployments, industrial companies should ensure that their cloud provider implements robust DDoS mitigation techniques and guarantees high availability through well-defined Service Level Agreements (SLAs).

Future research should focus on developing advanced DDoS mitigation strategies, including multi-cloud deployments that enhance resilience by distributing critical functions, as well as integrating effective security controls such as traffic filtering and re-deployment strategies [63,84].

**Centralized Control:** SDN, a key technique for virtualizing network infrastructure, introduces additional risks to the availability of an ICS due to its centralized control architecture. In SDN, the network is managed by a single unit called the SDN controller, which establishes end-to-end connections by installing flows in the transport network (switches). As a result, the controller becomes a single point of failure. If an attacker were to compromise the SDN controller via a DoS attack or otherwise, they could disrupt the entire network [85]. To mitigate this, *research should explore alternative, distributed SDN deployments* that avoid single points of failure while ensuring that additional latency is not introduced to the network, particularly in critical industrial data flows [86].

All the previously mentioned attack surfaces (virtualization, cloud deployments, and centralized control) require further research and development to ensure secure implementations under the 5G framework. Standardizing isolation mechanisms and secure programming frameworks for virtualized environments, enforcing strong cloud security controls, and investigating more resilient SDN architectures are critical steps to enhance the resilience of industrial networks in the 5G era.

*6.3. Fake base station attacks*

One of the most well-known threats to mobile networks is the so-called Fake Base Station (FBS), where an attacker impersonates a legitimate gNB to deceive the UE into connecting to it instead of the authentic base station. There are multiple methods to achieve this; however, the most common approach is signal overshadowing, where the FBS emits a stronger signal than the legitimate one. Consequently, the UE automatically connects to the gNB with the stronger signal quality [87]. Additionally, more stealthy techniques have been demonstrated, such as, e.g., SigUnder [88]. This attack can be utilized for various purposes, including data manipulation, UE tracking, and Denial of Service (DoS) [50,87,88].

5G introduced new security controls aimed at mitigating the effects of such attacks. The most notable example is the Subscriber Permanent Identifier (SUPI) encryption, which prevents the exposure of the UE identifier to attackers. However, these controls do not address the root cause of the issue—the lack of mutual authentication between the UE and the gNB [89]. As a result, attackers could still force UEs to connect to FBSs, leading to DoS attacks and complete loss of availability [89].

In commercial networks, multiple antennas are spread across a wide geographical area with strong transmission signals. In contrast, industrial networks often have fewer antennas, sometimes as few as a single antenna, with lower power, confining the signal within the company's boundaries. This makes it easier for an attacker to overshadow the signal, potentially disrupting the functionality of the entire industrial network.

The absence of adequate security measures may hinder the adoption of 5G in industrial settings that require very high availability. The 3GPP has acknowledged this issue and conducted research exploring several potential solutions, including certificate-based Public Key Infrastructure (PKI) and identity-based signature schemes [90]. However, these approaches introduced significant additional overhead to communication, making them impractical for low-latency scenarios.

Initial research shows that sub-millisecond overhead schemes are feasible. For example, Singla et al. [89] introduce a hierarchical identity-based signature scheme called Schnorr-HIBS, which eliminates the need for traditional certificates and allows for smaller signatures, thereby reducing communication overhead. In this scheme, a new entity called the core-PKG, which possesses a master key pair, delegates key management by providing key pairs derived from its master secret key to intermediate entities such as the AMF. The AMF, in turn, generates key pairs for the gNB based on its own secret key enabling them to sign messages. This interconnected structure ensures that all keys within the system are linked, allowing UEs to verify these signatures offline using the master public key stored in the SIM card. This scheme only adds roughly over $0.5$ ms of additional latency to the communication. Further research in this direction could potentially reduce this latency even more.

*6.4. Performance over security*

In time-critical industrial applications, latency and real-time operation are vital to ensure correct operation and safety [91]. However, cryptographic operations, especially those related to integrity protection, are known to introduce a notable amount of latency to communication [92]. Although the importance of UP security is critical (cf. Section 4.2.1), industrial companies are often required to make a trade-off between low latency and security. Typically, they sacrifice security in favor of performance, and thus not utilize the integrity protection of the UP. Especially, when safety requirements depend on the network quality (cf. Section 3.2).

To illustrate this issue, a recent study performed measurements to assess the overhead added by the integrity protection of the UP [93]. The results indicate that even in the low latency configuration of 5G, the additional overhead added to the round-trip time can render UP security unsuitable for industrial applications. Since these measurements did not account for additional security controls for UP data — such as securing the N3 interface — implementing these controls would introduce further latency, thereby increasing the overall cost of security. To address this issue, research on lightweight cryptography is necessary. More efficient cryptographic algorithms or message authentication code schemes should be studied and adopted in industrial 5G networks. Examples include AES-GCM, a scheme for authenticated encryption that speeds up processing by computing Message Authentication Code (MAC) and ciphertext in parallel, and is recommended in the technical report on URLLC security by 3GPP [94]. Another promising example is BP-MAC [92], which can be utilized for integrity protection and has been demonstrated to be faster than other lightweight integrity protection schemes, particularly for shorter messages, which are often encountered in industrial networks. Finally, as industrial devices often lack the computational power and/or hardware accelerators for traditional cryptographic schemes, or may have limited bandwidth (in mMTC/IIoT scenarios), alternative cryptographic schemes tailored for constrained environments [95], should be considered.

Moreover, Zeidler et al. investigated the utilization of TLS for communication between 5GC NFs [96]. They demonstrated that while the use of TLS typically results in an overhead of less than 1%, there are instances, such as after a system reboot, where the overhead increases to around 30%, which can be prohibitive in an industrial setting. As they suggest, alternative protocols such as IPSec and WireGuard should be investigated.

**Summary:** Despite its optimizations and potential in terms of security, 5G still faces significant challenges, especially concerning its secure integration into industrial networks. Further research is imperative to address crucial security aspects. The introduction of the wireless interface and new technologies poses substantial threats to system availability. Preventative measures are often lacking, with detection being the primary approach. Additionally, even when security controls could prevent attacks, they are often not implemented due to concerns about their impact on system performance.

## 7. Conclusion

5G stands out as the first and currently only wireless technology that meets the demands of industrial networks for reliable mobile connectivity and low latency. Consequently, an increasing number of companies start adopting 5G to leverage its multiple benefits without compromising any performance requirements. However, the advent of 5G significantly expands the attack surface of industrial networks by introducing multiple new components and a wireless interface. Therefore, the integration of 5G into industrial networks must prioritize security and safety as top concerns. In this paper, we provide a curated list of current state-of-the-art approaches essential for securely deploying and configuring a 5G network in an industrial setting, while also discussing promising opportunities to leverage existing technologies of 5G to further enhance security. Finally, we identify remaining challenges concerning 5G integration in industrial networks to fuel further research to address those. Overall, our work not only summarizes current research on securing industrial 5G networks and the exciting research challenges ahead but also provides a starting point to securely harness 5G's capabilities. This can unlock the potential for enhanced automation and mobility, inter-connection with vast cloud resources, and real-time operation, fulfilling the requirements of Industry 4.0 and IIoT.

**CRediT authorship contribution statement**

**Sotiris Michaelides:** Writing – original draft, Visualization, Investigation, Conceptualization. **Stefan Lenz:** Writing – review & editing, Investigation. **Thomas Vogt:** Writing – review & editing, Investigation. **Martin Henze:** Writing – review & editing, Supervision, Funding acquisition.

**Funding**

Funded by the Deutsche Forschungsgemeinschaft (DFG, German Research Foundation) under Germany's Excellence Strategy – EXC-2023 Internet of Production – 390621612 and by the German Federal Office for Information Security (BSI) under project funding reference numbers 01MO23016D (5G-Sierra) and 01MO24003B (CSII). The responsibility for the content of this publication lies with the authors.

**Declaration of competing interest**

The authors declare that they have no known competing financial interests or personal relationships that could have appeared to influence the work reported in this paper.

**Data availability**

No data was used for the research described in the article.

## References

[1] E.D. Knapp, Industrial Network Security: Securing Critical Infrastructure Networks for Smart Grid, SCADA, and Other Industrial Control Systems, 2024.
[2] M. Serror, S. Hack, M. Henze, M. Schuba, K. Wehrle, Challenges and opportunities in securing the industrial internet of things, IEEE Trans. Ind. Inform. (2021).
[3] A. Mahmood, L. Beltramelli, S. Fakhrul Abedin, S. Zeb, N.I. Mowla, S.A. Hassan, E. Sisinni, M. Gidlund, Industrial IoT in 5G-and-beyond networks: Vision, architecture, and design trends, IEEE Trans. Ind. Inform. (2022) 4122–4137, http://dx.doi.org/10.1109/TII.2021.3115697.
[4] J. Bodenhausen, C. Sorgatz, T. Vogt, K. Grafflage, S. Rötzel, M. Rademacher, M. Henze, Securing wireless communication in critical infrastructure: Challenges and opportunities, in: Proceedings of the 20th EAI International Conference on Mobile and Ubiquitous Systems: Computing, Networking and Services (MobiQuitous), 2023.
[5] M. Henze, M. Ortmann, T. Vogt, O. Ugus, K. Hermann, S. Nohr, Z. Lu, S. Michaelides, A. Massonet, R.H. Schmitt, Towards secure 5G infrastructures for production systems, in: Proceedings of the 22nd Conference on Applied Cryptography and Network Security (ACNS) – Poster Session, 2024, http://dx.doi.org/10.1007/978-3-031-61489-7_14.
[6] R. Langner, Stuxnet: Dissecting a cyberwarfare weapon, IEEE Secur. Priv. 9 (2011) 49–51, http://dx.doi.org/10.1109/MSP.2011.67.
[7] D.E. Whitehead, K. Owens, D. Gammel, J. Smith, Ukraine cyber-induced power outage: Analysis and practical mitigation strategies, in: 2017 70th Annual Conference for Protective Relay Engineers, CPRE, 2017, pp. 1–8, http://dx.doi.org/10.1109/CPRE.2017.8090056.
[8] T. Krause, R. Ernst, B. Klaer, I. Hacker, M. Henze, Cybersecurity in power grids: Challenges and opportunities, Sensors 21 (18) (2021) http://dx.doi.org/10.3390/s21186225.
[9] O. Lasierra, G. Garcia-Aviles, E. Municio, A. Skarmeta, X. Costa-Pérez, European 5G security in the wild: Reality versus expectations, in: ACM WiSec, 2023, http://dx.doi.org/10.1145/3558482.3581776.
[10] 5G-ACIA, 5G non-public networks for industrial scenarios, 2019, URL https://5g-acia.org/whitepapers/5g-non-public-networks-for-industrial-scenarios/, (Accessed 18 June 2024).

[11] A. Rostami, Private 5G networks for vertical industries: Deployment and operation models, in: IEEE 2nd 5G World Forum, 2019, http://dx.doi.org/10.1109/5GWF.2019.8911687.

[12] A. Aijaz, Private 5G: The future of industrial wireless, IEEE Ind. Electron. Mag. (2020) http://dx.doi.org/10.1109/MIE.2020.3004975.

[13] ENISA, Security in 5G specification, 2021, URL https://www.enisa.europa.eu/publications/security-in-5g-specifications, (Accessed 18 June 2024).

[14] FCC, Report on recommendations for identifying optional security features that can diminish the effectiveness of 5g security, 2021, URL https://www.fcc.gov/file/20606/download, (Accessed 18 June 2024).

[15] 5G-ACIA, Security Aspects of 5G for Industrial Networks, Tech. rep., 2021, URL https://5g-acia.org/whitepapers/security-aspects-of-5g-for-industrial-networks/, (Accessed 18 June 2024).

[16] 5G-ACIA, 5G for automation in industry, 2019, URL https://5g-acia.org/wp-content/uploads/2021/04/5G-ACIA_5G-for-Automation-in-Industry-.pdf, (Accessed 18 June 2024).

[17] 5G-ACIA, Key 5G use cases and requirements, 2019, URL https://5g-acia.org/wp-content/uploads/2021/04/Key_5G_Use_Cases_and_Requirements_DOWNLOAD.pdf, (Accessed 18 June 2024).

[18] S. Sullivan, A. Brighente, S.A.P. Kumar, M. Conti, 5G security challenges and solutions: A review by OSI layers, IEEE Access (2021) 116294–116314, http://dx.doi.org/10.1109/ACCESS.2021.3105396.

[19] A. Dutta, E. Hammad, 5G security challenges and opportunities: A system approach, in: 2020 IEEE 3rd 5G World Forum (5GWF), 2020, pp. 109–114, http://dx.doi.org/10.1109/5GWF49715.2020.9221122.

[20] J. Falco, F. Proctor, K. Stouffer, A. Wavering, IT security for industrial control systems, 2002, http://dx.doi.org/10.6028/NIST.IR.6859.

[21] E. Byres, The air gap: SCADA's enduring security myth, Commun. ACM (2013) 29–31, http://dx.doi.org/10.1145/2492007.2492018.

[22] J. Ke, Redundant wireless bridges: The reliability of wired networks with the cost savings of wireless, 2015, URL https://www.solutions.adm21.fr/redundancy-technology/Redundant_Wireless_Bridges_The_Reliability_of_Wired_Networks_with_the_Cost_Savings_of_Wireless.pdf, (Accessed 18 June 2024).

[23] D. Harutyunyan, D. Ginthoer, P. Buseck, S. Boje, A. Hohner, H.-J. Decker, N. Reider, G. Nemeth, S. Rácz, A. Vidács, G. Fehér, M. Maliosz, D. Patel, L. Grosjean, J. Sachs, E. Nacken, D. Hoheisel, in: D. Harutyunyan (Ed.), 5G-SMART 4.4 Report on Validation of 5G Use Cases in the Factory, Tech. rep., Bosch, ERI-HU, BME, ERI-DE, ERI-SE, 2022, URL https://5gsmart.eu/wp-content/uploads/5G-SMART-D4.4-v1.0.pdf, EU funded project: 5G-SMART.

[24] R. Kiesel, S. Schmitt, N. König, M. Brochhaus, T. Vollmer, K. Stichling, A. Mann, R.H. Schmitt, Techno-economic evaluation of 5G-NSA-NPN for networked control systems, Electronics http://dx.doi.org/10.3390/electronics11111736.

[25] 3GPP, TS 23.501 Version 17.5.0 Release 17, Tech. rep., 2022.

[26] P. Kehl, D. Lange, F.K. Maurer, G. Németh, D. Overbeck, S. Jung, N. König, R.H. Schmitt, Comparison of 5G enabled control loops for production, in: IEEE 31st IEEE Annual International Symposium on Personal, Indoor and Mobile Radio Communications, 2020.

[27] A. Seferagic, J. Famaey, E. De Poorter, J. Hoebeke, Survey on wireless technology trade-offs for the industrial Internet of Things, Sensors (2020) http://dx.doi.org/10.3390/s20020488.

[28] E. Oughton, W. Lehr, K. Katsaros, I. Selinis, D. Bubley, J. Kusuma, Revisiting wireless internet connectivity: 5G vs Wi-Fi 6, Telecommun. Policy (2021) http://dx.doi.org/10.1016/j.telpol.2021.102127.

[29] R. Maldonado, A. Karstensen, G. Pocovi, A.A. Esswie, C. Rosa, O. Alanen, M. Kasslin, T. Kolding, Comparing wi-fi 6 and 5G downlink performance for industrial IoT, IEEE Access 9 (2021) 86928–86937, http://dx.doi.org/10.1109/ACCESS.2021.3085896.

[30] W. Liang, J. Zhang, H. Shi, K. Wang, Q. Wang, M. Zheng, H. Yu, An experimental evaluation of WIA-FA and IEEE 802.11 networks for discrete manufacturing, IEEE Trans. Ind. Inform. (2021) http://dx.doi.org/10.1109/TII.2021.3051269.

[31] X. Jiang, M. Luvisotto, Z. Pang, C. Fischione, Reliable minimum cycle time of 5G NR based on data-driven channel characterization, IEEE Trans. Ind. Inform. (2021) http://dx.doi.org/10.1109/TII.2021.3052922.

[32] 5G Americas, Understanding 5G time critical services, 2022, URL https://www.5gamericas.org/wp-content/uploads/2022/08/Understanding-5G-Time-Critical-Services-Aug-2022.pdf, (Accessed 18 June 2024).

[33] IEEE 802.1, Time-sensitive networking (TSN) task group, 2020, URL https://1.ieee802.org/tsn/#TSN_Standards, (Accessed 18 June 2024).

[34] J. Farkas, L.L. Bello, C. Gunther, Time-sensitive networking standards, IEEE Commun. Stand. Mag. (2018) http://dx.doi.org/10.1109/MCOMSTD.2018.8412457.

[35] D. Cavalcanti, C. Cordeiro, M. Smith, A. Regev, WiFi TSN: Enabling deterministic wireless connectivity over 802.11, IEEE Commun. Stand. Mag. (2022) 22–29, http://dx.doi.org/10.1109/MCOMSTD.0002.2200039.

[36] 5G SMART - 5G for smart manufacturing, 2023, https://5gsmart.eu/, (Accessed 02 October 2023).

[37] J. Ansari, C. Andersson, P. de Bruin, J. Farkas, L. Grosjean, J. Sachs, J. Torsner, B. Varga, D. Harutyunyan, N. König, et al., Performance of 5G trials for industrial automation, Electronics (2022) http://dx.doi.org/10.3390/electronics11030412.

[38] Ö. Sen, D. van der Velde, P. Linnartz, I. Hacker, M. Henze, M. Andres, A. Ulbig, Investigating man-in-the-middle-based false data injection in a smart grid laboratory environment, in: Proceedings of 2021 IEEE PES Innovative Smart Grid Technologies Europe (ISGT-Europe), 2021, http://dx.doi.org/10.1109/ISGTEurope52324.2021.9640002.

[39] S. Nie, Y. Zhang, T. Wan, H. Duan, S. Li, Measuring the deployment of 5G security enhancement, in: Proceedings of the 15th ACM Conference on Security and Privacy in Wireless and Mobile Networks, WiSec '22, 2022, pp. 169–174, http://dx.doi.org/10.1145/3507657.3528559.

[40] A.A. Abd EL-Latif, B. Abd-El-Atty, S.E. Venegas-Andraca, W. Mazurczyk, Efficient quantum-based security protocols for information sharing and data protection in 5G networks, Future Gener. Comput. Syst. (2019) 893–906, http://dx.doi.org/10.1016/j.future.2019.05.053.

[41] E.J. Sacoto-Cabrera, L. Guijarro, J.R. Vidal, V. Pla, Economic feasibility of virtual operators in 5G via network slicing, Future Gener. Comput. Syst. (2020) 172–187, http://dx.doi.org/10.1016/j.future.2020.03.044.

[42] C.D. Alwis, P. Porambage, K. Dev, T.R. Gadekallu, M. Liyanage, A survey on network slicing security: Attacks, challenges, solutions and research directions, IEEE Commun. Surv. Tutor. (2023) http://dx.doi.org/10.1109/COMST.2023.3312349, 1–1.

[43] S. Köpsell, A. Ruzhanskiy, A. Hecker, D. Stachorra, N. Franchi, Open-RAN risk analysis, 2022, URL https://www.bsi.bund.de/SharedDocs/Downloads/EN/BSI/Publications/Studies/5G/5GRAN-Risk-Analysis.pdf?__blob=publicationFile&v=7, (Accessed 18 June 2024).

[44] S. Behrad, E. Bertin, S. Tuffin, N. Crespi, A new scalable authentication and access control mechanism for 5G-based IoT, Future Gener. Comput. Syst. (2020) 46–61, http://dx.doi.org/10.1016/j.future.2020.02.014.

[45] I. Ahmad, T. Kumar, M. Liyanage, J. Okwuibe, M. Ylianttila, A. Gurtov, Overview of 5G security challenges and solutions, IEEE Commun. Stand. Mag. 2 (2018) 36–43, http://dx.doi.org/10.1109/MCOMSTD.2018.1700063.

[46] 3GPP, TS 33.501 version 17.5.0 Release 17, Tech. rep., 2022.

[47] M. Dahlmanns, J. Lohmöller, J. Pennekamp, J. Bodenhausen, K. Wehrle, M. Henze, Missed opportunities: Measuring the untapped TLS support in the industrial internet of things, in: Proceedings of the 17th ACM ASIA Conference on Computer and Communications Security, ASIA CCS, 2022, http://dx.doi.org/10.1145/3488932.3497762.

[48] R. Security, 5G SCAS evaluation access and mobility management function AMF 3GPP TS 33.512 version 16.3.0 Release 16), 2022, URL https://radix-security.com/files/2022/sample_report/sample_report.pdf, (Accessed 18 June 2024).

[49] D.T. Hoang, C. Park, M. Son, T. Oh, S. Bae, J. Ahn, B. Oh, Y. Kim, LTESniffer: An open-source LTE downlink/uplink eavesdropper, in: 16th ACM Conference on Security and Privacy in Wireless and Mobile Networks (WiSec '23), 2023.

[50] D. Rupprecht, K. Kohls, T. Holz, C. Pöpper, Breaking LTE on layer two, in: 2019 IEEE Symposium on Security and Privacy, 2019, http://dx.doi.org/10.1109/SP.2019.00006.

[51] M. Akon, T. Yang, Y. Dong, S.R. Hussain, Formal analysis of access control mechanism of 5G core network, in: Proceedings of the 2023 ACM SIGSAC Conference on Computer and Communications Security, CCS '23, 2023, pp. 666–680, http://dx.doi.org/10.1145/3576915.3623113.

[52] W. Yan, R. Chan, T. Truong-Huu, 5G Core Security: An Insider Threat Vulnerability Assessment, 2024.

[53] S. Park, A. Shaik, R. Borgaonkar, J.-P. Seifert, Anatomy of commercial IMSI catchers and detectors, in: Proceedings of the 18th ACM Workshop on Privacy in the Electronic Society, WPES '19, Association for Computing Machinery, 2019, pp. 74–86, http://dx.doi.org/10.1145/3338498.3358649.

[54] S.F. Mjølsnes, R.F. Olimid, Easy 4G/LTE IMSI catchers for non-programmers, 2017, arXiv:1702.04434.

[55] M. Chlosta, D. Rupprecht, C. Pöpper, T. Holz, 5G SUCI-Catchers: Still catching them all? in: ACM WiSec, WiSec '21, 2021, http://dx.doi.org/10.1145/3448300.3467826.

[56] X. Huang, T. Yoshizawa, S.B.M. Baskaran, Authentication mechanisms in the 5G system, J. ICT Stand. (2021) 61–78, http://dx.doi.org/10.13052/jicts2245-800X.921.

[57] J. Vollbrecht, J.D. Carlson, L. Blunk, D.B.D. Aboba, H. Levkowetz, Extensible authentication protocol (EAP), 2004, http://dx.doi.org/10.17487/RFC3748.

[58] P. De Vaere, A. Tulimiero, A. Perrig, Hopper: Per-device nano segmentation for the industrial IoT, in: Proceedings of the 2022 ACM on Asia Conference on Computer and Communications Security, in: ASIA CCS '22, Association for Computing Machinery, New York, NY, USA, 2022, pp. 279–293, http://dx.doi.org/10.1145/3488932.3501277.

[59] S. Wijethilaka, M. Liyanage, Realizing internet of things with network slicing: Opportunities and challenges, in: IEEE 18th Annual Consumer Communications and Networking Conference, 2021, pp. 1–6, http://dx.doi.org/10.1109/CCNC49032.2021.9369637.

[60] L. Bader, M. Serror, O. Lamberts, Ö. Sen, D. van der Velde, I. Hacker, J. Filter, E. Padilla, M. Henze, Comprehensively analyzing the impact of cyberattacks on power grids, in: Proceedings of the 2023 IEEE 8th European Symposium on Security and Privacy (EuroS&P), 2023, http://dx.doi.org/10.1109/EuroSP57164.2023.00066.

[61] R. Hussain, F. Hussain, S. Zeadally, Integration of VANET and 5G security: A review of design and implementation issues, Future Gener. Comput. Syst. (2019) 843–864, http://dx.doi.org/10.1016/j.future.2019.07.006.
[62] ENISA, NFV Security in 5G - Challenges and Best Practices, Tech. rep., 2022, URL https://www.enisa.europa.eu/publications/nfv-security-in-5g-challenges-and-best-practices, (Accessed 18 June 2024).
[63] M.J. Shayegan, A. Damghanian, A review of methods to prevent DDOS attacks using NFV and SDN, in: 2023 9th International Conference on Web Research, ICWR, 2023, pp. 346–355, http://dx.doi.org/10.1109/ICWR57742.2023.10139112.
[64] G.W. De Oliveira, M. Nogueira, A.L.d. Santos, D.M. Batista, Intelligent VNF placement to mitigate ddos attacks on industrial IoT, IEEE Trans. Netw. Serv. Manag. (2023) 1319–1331, http://dx.doi.org/10.1109/TNSM.2023.3274364.
[65] L. Zhou, H. Guo, G. Deng, A fog computing based approach to ddos mitigation in iIoT systems, Comput. Secur. (2019) 51–62, http://dx.doi.org/10.1016/j.cose.2019.04.017.
[66] S. Erokhin, A. Petukhov, P. Pilyugin, Critical Information Infrastructures Monitoring Based on Software-Defined Networks, 2019, URL https://api.semanticscholar.org/CorpusID:201900161.
[67] T. Garfinkel, M. Rosenblum, A virtual machine introspection based architecture for intrusion detection, in: Network and Distributed System Security Symposium, 2003, URL https://api.semanticscholar.org/CorpusID:6136159.
[68] W. O-RAN Alliance, O-RAN Architecture Description 12.0, Tech. rep., 2023.
[69] H. Wen, P. Porras, V. Yegneswaran, A. Gehani, Z. Lin, 5G-spector: An O-RAN compliant layer-3 cellular attack detection service, in: Proceedings of the 31st Annual Network and Distributed System Security Symposium, NDSS'24, 2024.
[70] D. Karnwal, Anomaly detection use case, 2021, (Accessed 18 June 2024).
[71] P. Kryszkiewicz, M. Hoffmann, Open RAN for detection of a jamming attack in a 5G network, in: 2023 IEEE 97th Vehicular Technology Conference (VTC2023-Spring), 2023, pp. 1–2, http://dx.doi.org/10.1109/VTC2023-Spring57618.2023.10201067.
[72] K. Wolsing, E. Wagner, A. Saillard, M. Henze, IPAL: Breaking up silos of protocol-dependent and domain-specific industrial intrusion detection systems, in: Proceedings of the 25th International Symposium on Research in Attacks, Intrusions and Defenses, RAID, 2022, http://dx.doi.org/10.1145/3545948.3545968.
[73] S. Maesschalck, V. Giotsas, B. Green, N. Race, Don't get stung, cover your ICS in honey: How do honeypots fit within industrial control system security, Comput. Secur. (2022) 102598, http://dx.doi.org/10.1016/j.cose.2021.102598.
[74] M. Lichtman, R. Rao, V. Marojevic, J. Reed, R.P. Jover, 5G NR jamming, spoofing, and sniffing: Threat assessment and mitigation, in: 2018 IEEE ICC Workshops, 2018, pp. 1–6, http://dx.doi.org/10.1109/ICCW.2018.8403769.
[75] A. Cetinkaya, H. Ishii, T. Hayakawa, Effects of jamming attacks on wireless networked control systems under disturbance, IEEE Trans. Autom. Control (2023) 1223–1230, http://dx.doi.org/10.1109/TAC.2022.3153275.
[76] Y. Arjoune, S. Faruque, Smart jamming attacks in 5G new radio: A review, in: IEEE 10th Annual Computing and Communication Workshop and Conference, 2020, pp. 1010–1015, http://dx.doi.org/10.1109/CCWC47524.2020.9031175.
[77] S. Barros, J. Bazzo, R. Takaki, D. Carrillo, J. Seki, LTE jamming mitigation based on frequency hopping strategies, in: 2016 8th IEEE Latin-American Conference on Communications, LATINCOM, 2016, pp. 1–6, http://dx.doi.org/10.1109/LATINCOM.2016.7811609.
[78] A. Liu, P. Ning, H. Dai, Y. Liu, C. Wang, Defending DSSS-based broadcast communication against insider jammers via delayed seed-disclosure, in: ACM ACSAC, ACSAC '10, New York, NY, USA, 2010, pp. 367–376, http://dx.doi.org/10.1145/1920261.1920315.
[79] C.-Y. Lin, S. Nadjm-Tehrani, M. Asplund, Timing-based anomaly detection in SCADA networks, in: Critical Information Infrastructures Security, 2018, pp. 48–59.
[80] ENISA, Security Aspects of Virtualization, Tech. rep., 2017, URL https://www.enisa.europa.eu/publications/security-aspects-of-virtualization, (Accessed 18 June 2024).
[81] O. AbdElRahem, A.M. Bahaa-Eldin, A. Taha, Virtualization security: A survey, in: 2016 11th International Conference on Computer Engineering & Systems, ICCES, 2016, pp. 32–40, http://dx.doi.org/10.1109/ICCES.2016.7821971.
[82] R.F. Olimid, G. Nencioni, 5G network slicing: A security overview, IEEE Access (2020) 99999–100009, http://dx.doi.org/10.1109/ACCESS.2020.2997702.
[83] M. Jensen, J. Schwenk, N. Gruschka, L.L. Iacono, On technical security issues in cloud computing, in: 2009 IEEE International Conference on Cloud Computing, 2009, pp. 109–116, http://dx.doi.org/10.1109/CLOUD.2009.60.
[84] B.B. Gupta, O. Badve, Taxonomy of DoS and ddos attacks and desirable defense mechanism in a cloud computing environment, Neural Comput. Appl. (2017) http://dx.doi.org/10.1007/s00521-016-2317-5.
[85] J.C. Correa Chica, J.C. Imbachi, J.F. Botero Vega, Security in SDN: A comprehensive survey, J. Netw. Comput. Appl. (2020) 102595, http://dx.doi.org/10.1016/j.jnca.2020.102595.
[86] D. Ryait, M. Sharma, To eliminate the threat of a single point of failure in the SDN by using the multiple controllers, Int. J. Recent Technol. Eng. (IJRTE) (2020) 234–241, http://dx.doi.org/10.35940/ijrte.B3433.079220.
[87] H. Yang, S. Bae, M. Son, H. Kim, S.M. Kim, Y. Kim, Hiding in plain signal: physical signal overshadowing attack on LTE, in: USENIX SEC, SEC '19, 2019, pp. 55–72.
[88] N. Ludant, G. Noubir, SigUnder: a stealthy 5G low power attack and defenses, in: ACM WiSec, WiSec '21, 2021, pp. 250–260, http://dx.doi.org/10.1145/3448300.3467817.
[89] A. Singla, R. Behnia, S.R. Hussain, A. Yavuz, E. Bertino, Look before you leap: Secure connection bootstrapping for 5G networks to defend against fake base-stations, in: ASIA CCS, ASIA CCS '21, 2021, pp. 501–515, http://dx.doi.org/10.1145/3433210.3453082.
[90] 3GPP, 3GPP TR 33.809 Version 0.3.0 Release 16, Tech. rep., 2019.
[91] J. Hiller, M. Henze, M. Serror, E. Wagner, J.N. Richter, K. Wehrle, Secure low latency communication for constrained industrial IoT scenarios, in: Proceedings of the 43rd IEEE Conference on Local Computer Networks, LCN, 2018, http://dx.doi.org/10.1109/LCN.2018.8638027.
[92] E. Wagner, M. Serror, K. Wehrle, M. Henze, BP-MAC: Fast Authentication for Short Messages, WiSec '22, 2022.
[93] T. Heijligenberg, G. Knips, C. Böhm, D. Rupprecht, K. Kohls, BigMac: Performance overhead of user plane integrity protection in 5g networks, in: ACM WiSec, WiSec '23, 2023, http://dx.doi.org/10.1145/3558482.3581777.
[94] 3GPP, TR 33.825 version 16.0.1 Release 16, Tech. rep., 2019.
[95] Y. Zhong, J. Gu, Lightweight block ciphers for resource-constrained environments: A comprehensive survey, Future Gener. Comput. Syst. (2024) 288–302, http://dx.doi.org/10.1016/j.future.2024.03.054.
[96] O. Zeidler, J. Sturm, D. Fraunholz, W. Kellerer, Performance evaluation of transport layer security in the 5G core control plane, in: Proceedings of the 17th ACM Conference on Security and Privacy in Wireless and Mobile Networks, WiSec '24, 2024, pp. 78–88, http://dx.doi.org/10.1145/3643833.3656140.

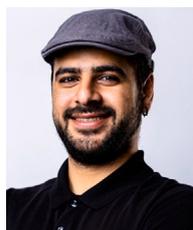

**Sotiris Michaelides** is a Ph.D. student and researcher in the Security and Privacy in Industrial Cooperation (SPICe) group at RWTH Aachen University in Aachen, Germany. He received his bachelor's degree in Computer Science in 2019 from the University of Cyprus in Cyprus, and then pursued a master's degree in Cyber Security at Radboud University in the Netherlands. His research interests lie within the area of mobile networks (5G) and emerging technologies.

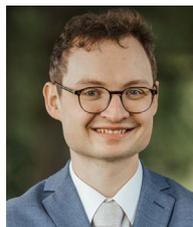

**Stefan Lenz** is a researcher in the Security and Privacy in Industrial Cooperation (SPICe) group at RWTH Aachen University in Aachen, Germany. He received his M.Sc. in Computer Science from RWTH Aachen University in 2024. His interests are in intrusion detection systems and their interaction with wireless communication systems.

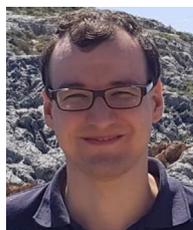

**Thomas Vogt** is a researcher in the Security and Privacy in Industrial Cooperation (SPICe) group at RWTH Aachen University in Aachen, Germany. He received his B.Sc. and M.Sc. in Computer Science from the same university. His research focuses on the operation and security of (radio) networks, involving two research projects.

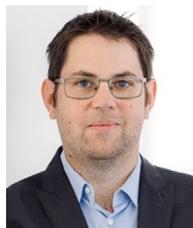

**Martin Henze** is a tenure-track assistant professor in the Department of Computer Science at RWTH Aachen University in Aachen, Germany, where he leads the Security and Privacy in Industrial Cooperation (SPICe) group. His research focuses on security and privacy in industrial networks and data sharing, particularly in the energy and production sectors. Additionally, he co-leads the Secure Production & Energy Networks research group within the Cyber Analysis & Defense department at the Fraunhofer Institute FKIE in Bonn, Germany. He earned both his Diploma (equiv. M.Sc.) and Ph.D. (Dr. rer. nat., with distinction) in Computer Science from RWTH Aachen University.